# Non-thermal Atmospheric Pressure Plasma Combined with Benzoyl Peroxide to Enhance the Inhibition of Bacteria

*Tzu-Hsuan Chen[1], Chao-Yu Chen[1], Yuan-Min Lin[2], Yun-Chien Cheng[3]*

---

[1] Department of Mechanical Engineering, National Yang Ming Chiao Tung University, Hsinchu 300093, Taiwan

[2] Department of Dentistry, College of Dentistry, National Yang Ming Chiao Tung University, Taipei 112304, Taiwan.

[3] Department of Electrical Engineering, National Taiwan University, Taipei 106319, Taiwan

Corresponding author: Yun-Chien Cheng, Yuan-Min Lin

E-mail: yunchien@ntu.edu.tw, ymlin@nycu.edu.tw

## Abstract

This study aims to verify whether combining non-thermal equilibrium atmospheric pressure plasma with benzoyl peroxide (BPO) can enhance the inhibition of bacteria. This work involves not only adjusting the parameters of the plasma device, such as increasing the power supply voltage, modifying the working gas flow rate, as well as incorporating a specific proportion of oxygen to the working gas, but also aims to combine plasma with BPO to inactivate bacteria. An optical emission spectroscopy was utilized to determine the optimal working gas flow rate and oxygen mixing ratio for the experimental setup. Additionally, surface temperature and ultraviolet (UV) emission measurements were also performed to characterize the physical properties of the plasma treatment. Although the precise chemical mechanisms between plasma and BPO have yet to be fully elucidated, bacterial inactivation experiments on *Escherichia coli* demonstrated that the addition of BPO indeed enhanced the sterilization efficacy of plasma treatment.

# I. Introduction

Plasma has emerged as a promising antibacterial agent. Plasma is a partially ionized gas composed of electrons, ions, and neutral particles.[1] The reactive oxygen species (ROS) generated by plasma exhibit strong oxidative properties capable of destroying bacteria without leaving residues in the body,[2] thereby avoiding the risk of antibiotic resistance.[3] Moreover, cold atmospheric plasma (CAP) operates at near-room temperature and generally maintains treated tissue temperatures under thermal limitation, supporting its feasibility for biomedical applications without causing significant thermal injury to biological tissues.[4] Most notably, its gaseous nature allows for deeper penetration into cavity or porous structures including dentinal tubules.[5] Consequently, plasma is considered a potential alternative to ensure complete bacterial eradication in biomedical field.

Effective bacterial elimination is essential for successful infection control. Complete eradication of bacteria is a critical step, as residual bacteria may continue to proliferate and damage the affected tissues, leading to treatment failure or even secondary infections that require more complex therapeutic interventions.

Therefore, enhancing the antibacterial efficacy of plasma is a crucial issue. Plasma reveals broad-effect on antibacterial activity might be primarily through the action of hydroxyl radicals ($\cdot OH$) and peroxynitrite ($ONOO^-$) present in the plasma.[6] Common methods to enhance the concentration of ROS in plasma include increasing the voltage of the plasma power supply,[7] adjusting the flow rate of the working gas,[8] and adding specific proportions of oxygen to the working gas.[9] In addition to these methods, which involve altering the parameters of the plasma device itself, this study aims to explore a novel approach by combining plasma with chemical substances in an attempt to further improve its antibacterial effect.

Benzoyl peroxide (BPO) is a common organic peroxide with strong oxidative properties that can cause cellular damage. Because of its antibacterial activity, it is widely used in dermatology, especially for the treatment of acne. [10] It is also used as the initiator of polymerization in dental resins.

The benzoyloxyl radical (BO radical) formed upon BPO decomposition can react with biomolecules, leading to cellular injury. Therefore, promoting the decomposition of BPO to generate BO radicals is a key factor in enhancing the bactericidal effect.[11]

BPO can decompose through heating, exposure to light, or interaction with metal ions or free radicals.[12-15] Heating requires a temperature of 75°C, which exceeds the tolerance range of the human body; therefore, this study focuses on the photodecomposition and radical-induced decomposition of BPO. Light with wavelengths between 365 and 480 nm can induce photolysis of BPO, generating BO radicals and atomic hydrogen. If further irradiated with ultraviolet light of wavelengths between 236 and 300 nm, the BO radicals are converted into phenyl radicals (Ph radicals) and release carbon dioxide.[13]

According to the mechanism proposed by Denney's team,[14] the α-ethoxyethyl radical ($CH_3CHOC_2H_5\cdot$) reacts with BPO by attacking the oxygen atom of the peroxide bond, leading to the decomposition of BPO. This reaction generates 1-ethoxyethyl benzoate and a BO radical.

$$(C_6H_5COO)_2 + CH_3\dot{C}HOC_2H_5 \rightarrow \left[ C_6H_5\overset{O}{\overset{\|}{C}}\underset{\vdots \atop R}{O}\dot{-}\!\!-OC\overset{O}{\overset{\|}{}}C_6H_5 \right]^{\ddagger}$$

Based on the principles of photolysis and radical-induced decomposition described in the previous literature, we expect that plasma-generated ultraviolet light and free radicals will facilitate the decomposition of BPO, generating highly reactive radicals that enhance the bactericidal effect. This approach may offer a new potential strategy for the treatment of bacteria-infected surfaces in both dermatological and dental care. In dermatology, BPO is already widely used for acne treatment due to its oxidative antibacterial properties, while cold plasma offers additional advantages such as non-thermal operation and the generation of ROS. Their combined use may enable improved penetration and more effective disruption of bacterial biofilms on the skin, thereby revealing the potential to reduce treatment duration and minimizing the required dosage of BPO and lowers the risk of skin irritation. In dentistry, where complete eradication of bacteria within complex structures such as dentinal tubules is critical, the gaseous nature of plasma allows deeper access to confined regions that are difficult to reach. The addition of BPO may further amplify antibacterial action through complementary radical-based mechanisms, improving disinfection outcomes in procedures such as cavity treatment or root canal therapy.

## II. Materials and Methods

### A. Atmospheric Pressure Plasma Experimental Setup

The experimental setup for this study is illustrated in Figure 1. An alumina tube ($Al_2O_3$ tube, Tochance , Taoyuan City, Taiwan) was selected as the dielectric material. The inner and outer diameters of the alumina tube were 3 mm and 5 mm, respectively. A stainless steel rod (diameter: 1 mm, Ultimate Materials, Hsinchu County, Taiwan) was used as the high-voltage electrode, positioned between the dielectric materials. The grounded electrode was an aluminum tape (length: 5 cm) attached to the surface of the dielectric material, with a distance of 5 mm from the outlet of the alumina tube.

In this study, a power supply (HVPS-S10, Naiding Technology, Tainan City, Taiwan) was used to provide an AC voltage of 4.3 kV (with Vmax ranging from 4.25 to 4.35 kV) at a frequency of 21.55 kHz. A high-voltage probe (P6015A, Tektronix, Oregon, United States) was employed to measure the plasma voltage. A digital oscilloscope (DSOX3024T, Keysight Technologies, California, United States) was used to record and observe the plasma voltage. Argon gas was used as the working gas, and its flow rate was controlled at 4 Standard Liters per Minute (SLM) using a flowmeter (0–20 SLM, Yongxin Technology, Hsinchu City, Taiwan).

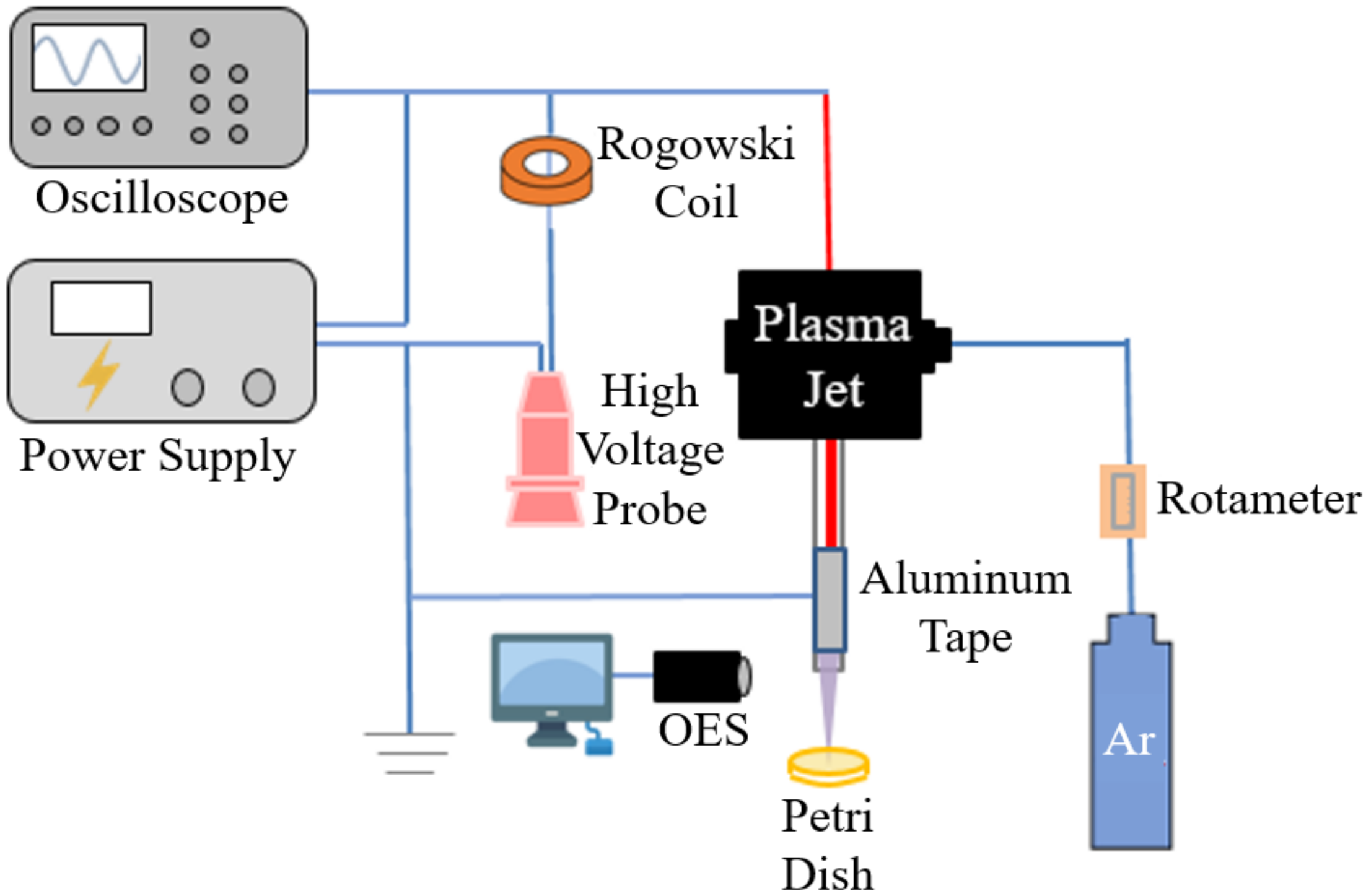

**FIG. 1:** Schematic diagram of the experimental setup.

## B. Measurement Methods and Instruments

### 1. Optical Emission Spectroscopy, OES

The Optical Emission Spectrometer (OES) is connected to the detection lens and spectrometer via optical fibers, collecting and analyzing the spectral signals emitted by the plasma. The position of the optical collection lens of the OES in this study is shown in Figure 1. The lens is placed horizontally and focused at a distance of 0 mm from the outlet of the alumina tube, where the light emission from the plasma jet at the tube's exit is transmitted through an optical fiber to the spectrometer and measured in a wide wavelength range.

For UV measurement, only wavelengths below 400 nm were considered as UV light. Background correction was applied to reduce ambient light interference.

**2. Forward Looking Infrared, FLIR**

We used an infrared thermal camera (FLIR ONE® Edge Pro, Teledyne Technologies Incorporated, California, USA) to measure the temperature.

In this study, ceramic materials were used to simulate the tooth surface. The infrared thermal imaging camera was employed to measure the temperature rise at the plasma application points on the ceramic surface during plasma treatment, with measurements taken continuously for five minutes.

**3. Chemical Titration**

A chemical titration method was employed to quantify the amount of BPO. The procedure involved initially dissolving 100 mg of BPO in 10 mL of hydroxyethyl methacrylate (HEMA, Emperor Chemical, Taipei City, Taiwan). Subsequently, 5 mL of 1 M potassium iodide (KI) solution (Uni-Onward, New Taipei City, Taiwan) was added to the mixture, which was then thoroughly mixed and allowed to stand for 15 minutes. The iodide ions react with BPO to produce molecular iodine ($I_2$), resulting in a dark brown coloration. As iodine is sensitive to light and may volatilize upon exposure to air, the entire process was conducted under light-protected and sealed conditions to ensure the accuracy of the results. The released iodine was then titrated using a 1 M sodium thiosulfate ($Na_2S_2O_3$) solution (Uni-Onward). The endpoint of the titration was reached when the solution changed from dark brown to colorless. Finally, the purity percentage (X) of the BPO sample was calculated according to Eq. (1), based on the volume of sodium thiosulfate solution used, enabling the quantitative analysis of BPO.[16]

$$X = (242.23 \times C \times V)/(2 \times m \times 1000) \times 100\ \% \qquad (1)$$

In equation (1), 242.23 represents the molar mass of BPO (g); C is the molar concentration of the sodium thiosulfate standard solution (mol/L); V is the volume of sodium thiosulfate solution consumed during titration (mL); and m is the mass of the BPO sample (g).

The BPO content was measured both before and after 5 minutes of continuous plasma treatment to determine whether plasma application leads to the decomposition of BPO and a corresponding reduction in its quantity.

**4. Antibacterial Assay**

The bactericidal efficacy of plasma alone and plasma combined with BPO was evaluated by measuring the zones of inhibition for *Escherichia coli* (AY-201911006, Food Industry Research and Development Institute, Hsinchu City, Taiwan). All other experimental conditions, including plasma source voltage, working gas flow rate, treatment distance, and treatment duration, were kept constant. By comparing the size of the zone of inhibition under each condition, the effect of BPO addition on the inactivation of bacteria was assessed.

LB broth (Bio Basic, Toronto, Canada) and agar powder (Bio Basic, Toronto, Canada) were first mixed to prepare an agarose gel medium containing LB broth nutrients, following sterilization in an autoclave (TM-326, Tomin medical equipment, New Taipei City, Taiwan). A 100 µL *E. coli* ($1.18 \times 10^9$ CFU/mL) culture were spread onto the surface of the agarose gel medium using glass beads. After the bacterial suspension dried, the subsequent steps were performed. For plates without BPO, plasma sterilization was applied directly. For plates with BPO, the BPO solution was spread onto the plate's surface using glass beads, and after the BPO solution dried, plasma sterilization was applied. To avoid the occurrence of arcing due to excessive proximity of the plasma outlet to the solution surface, a treatment distance of 17 mm was adopted during plasma application. The zone of inhibition was measured using Image J software.

## III. Results and Discussion

### A. Optical Emission Spectroscopy, OES

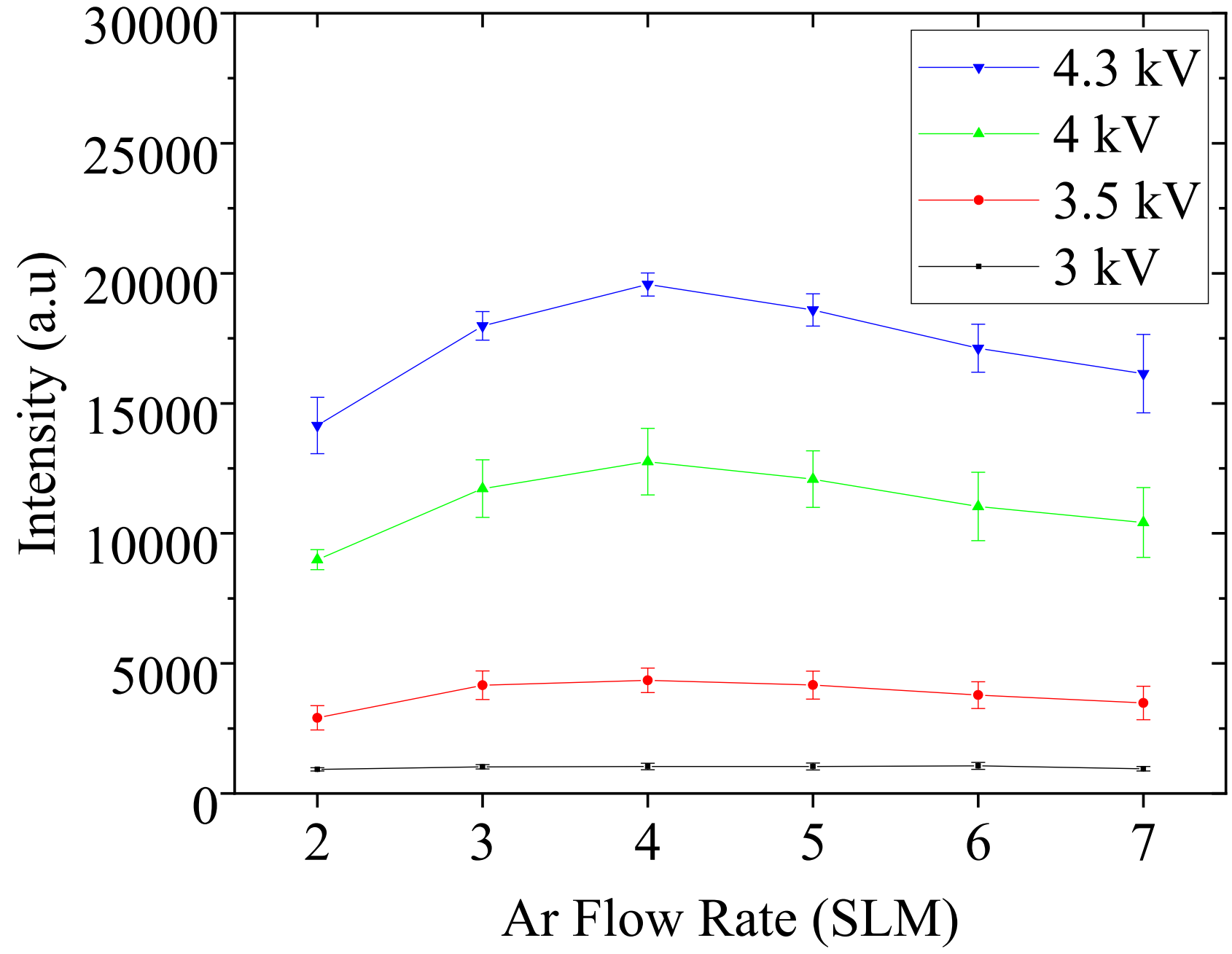

**FIG. 2:** Emission intensity at the 309-nm wavelength under different voltages and argon gas flow rates.

The voltage for this measurement ranged from 3 kV to 4.3 kV, with a frequency of 21.55 kHz. The argon flow rate ranged from 2 to 7 SLM, and the integration time for the OES measurement was 1300 μs. Figure 2 shows the measurement results under different voltages and argon flow rates. The intensity at the 309 nm wavelength exhibits an initial increase followed by a decrease as the argon flow rate increases.

According to the OES measurement results, the signal intensities observed at argon flow rates of 2 and 3 SLM were relatively low. This is presumed to be due to insufficient gas flow to effectively carry the plasma-generated radicals out of the nozzle.[17] When the flow rate exceeded 4 SLM, the 309 nm OES intensity gradually decreased with increasing flow rate. The influence of gas flow rate on the production of reactive oxygen species (ROS) in plasma is multifaceted and complex, and it can vary depending on experimental conditions. We speculate that the reason behind our experimental results may be that higher gas flow rates lead to increased gas particle density, which raises the probability of collisions between particles. As a result, the acceleration distance available to each electron becomes insufficient, reducing the overall dissociation rate and, consequently, the production of radicals. Additionally, higher gas flow rates tend to lower the overall plasma temperature, which reduces the chance of generating high-energy particles and thus decreases the efficiency of ROS production.

Moreover, it was confirmed that increasing the applied voltage enhances plasma intensity. However, since the ultimate application of this technology is within the human oral cavity, safety concerns must be considered. To prevent the occurrence of arc discharge, the applied voltage was limited to 4.3 kV.

At 4.3 kV and a flow rate of 4 SLM, the emission intensity at 309 nm was found to be the highest. Therefore, the subsequent experiments were conducted under these conditions, using an argon flow rate of 4 SLM.

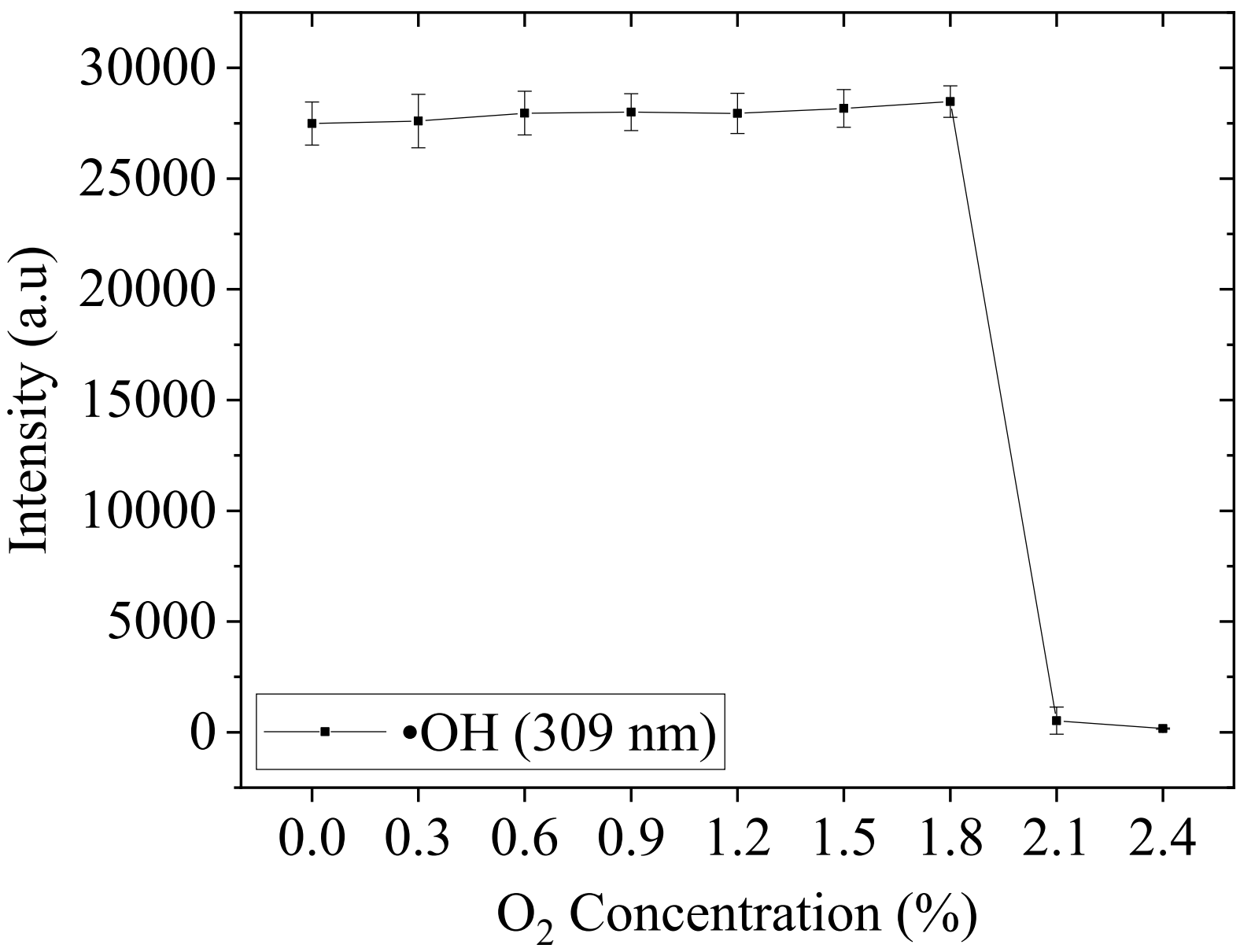

**FIG. 3:** Emission intensity at 309 nm under different oxygen addition ratios.

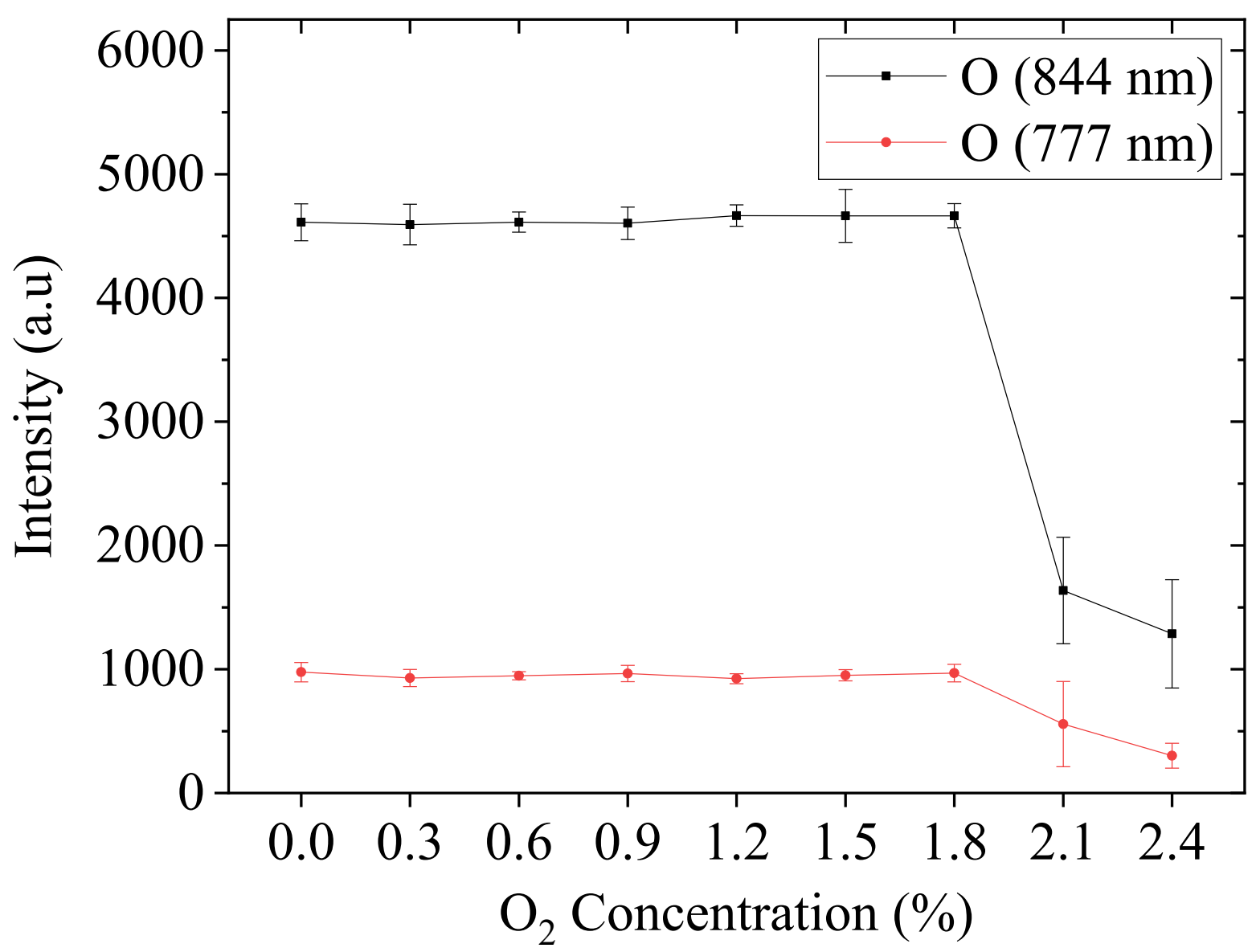

**FIG. 4:** Emission intensities at 777 nm and 844 nm under different oxygen addition ratios.

To possibly enhance the concentrations of OH and O radicals, oxygen is introduced into the working gas. Figures 3 and 4 illustrate the emission intensity variations at wavelengths of 309 nm, 777 nm, and 844 nm with different oxygen addition ratios ranging from 0% to 2.4% into the 4-slm argon plasma working gas. The integration time for OES was set at 1500 μs.

Despite previous reports suggesting that the addition of a certain proportion of oxygen to the plasma working gas can enhance the generation of reactive oxygen species, our experimental results demonstrated that emission intensities at 309 nm (OH radicals), 777 nm, and 844 nm (atomic oxygen) did not exhibit significant changes when the oxygen concentration was below 2%. Moreover, when the oxygen percentage exceeded 2%, a noticeable decrease in plasma density occurred, which is likely due to the electron-scavenging property of oxygen that reduces the population of free electrons in the plasma.[18] Consequently, pure argon was selected as the working gas for subsequent experiments.

### B. Forward Looking Infrared, FLIR

To monitor the surface temperature during plasma treatment, this experiment utilized ceramic materials to simulate tooth surfaces, assessing the temperature under plasma exposure. We measured the temperature at the plasma application point on the ceramic surface using an infrared thermal imaging device. Based on the room temperature, the temperature could differ. From one measurement, after 5 minutes of plasma exposure, the ceramic surface temperature increased from 18.4°C to 27.3°C (Fig. 4 in Appendix).

### C. Ultraviolet Measurement

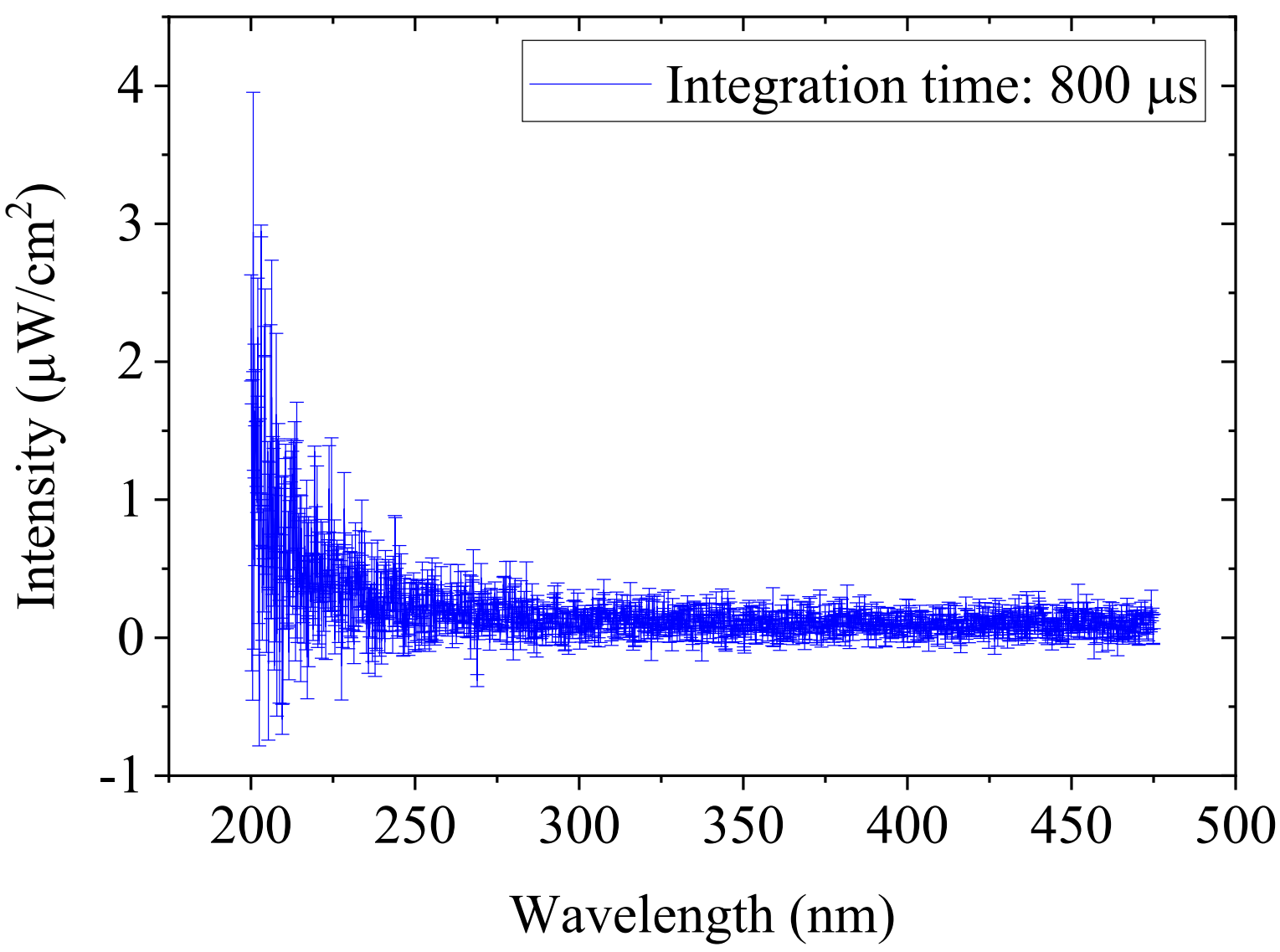

**FIG. 5:** UV measurement of plasma emission.

Figure 5 shows the measured UV intensity via OES with an integration time of 800 µs to illustrate plasma UV emission. The optical collection lens was aligned to the plasma jet exit, and the wavelengths below 475 nm were recorded.

Previous studies have shown that the initial homolytic cleavage of the peroxide O–O bond can be induced by irradiation in the 365–480 nm range, while subsequent decomposition of the resulting benzoyloxy radicals, including decarboxylation to phenyl radicals and $CO_2$, occurs primarily under shorter wavelengths in the 236–300 nm range.[13] In the present study, the plasma emitted 57.99 ± 22.03 $\mu W/cm^2$ in the 365–480 nm region and 57.67 ± 22.57 $\mu W/cm^2$ in the 236–300 nm region. The estimated photon flux was on the order of $10^{12}$ photons/$cm^2\cdot s$. Considering the reported low quantum yield of benzoyl peroxide photolysis (~$10^{-5}$),[13] the resulting radical generation via UV alone was expected to be limited, corresponding to a small fraction of total BPO molecules. This suggests that direct photolysis was not the dominant decomposition pathway under the present conditions.

### D. Chemical Titration

We aim to confirm whether BPO has been decomposed by quantitatively measuring its content. Theoretically, a decrease in BPO concentration would suggest that BPO has undergone decomposition.

**TABLE 1:** Measurement of Pure BPO Content Using Chemical Titration Method

| Substance | Treatment | Result |
|---|---|---|
| **Impure BPO 100 mg** | X | Pure BPO 79.1 (±1.4) mg |
| **Impure BPO 100 mg** | Apply plasma treatment for five minutes | Pure BPO 78.7 (±1.2) mg |

The quantification result of 100 mg of non-pure BPO shows that it contains 79.1 (±1.4) mg of pure BPO. After 5 minutes of continuous plasma treatment, the quantification result of 100 mg of non-pure BPO shows that it contains 78.7 (±1.2) mg of pure BPO.

Although the quantification results from the chemical titration method indicate that plasma treatment did not cause a decrease in BPO content, suggesting that plasma may not be able to induce BPO decomposition, we cannot rule out the possibility that BPO, after decomposing into free radicals, might potentially revert back to BPO, which could explain why there was no decrease in BPO content. Therefore, based on the current experimental results, we are unable to determine conclusively whether plasma directly decomposes BPO. However, if such decomposition occurs, its extent is likely limited under the present experimental conditions.

### E. Antibacterial Assay

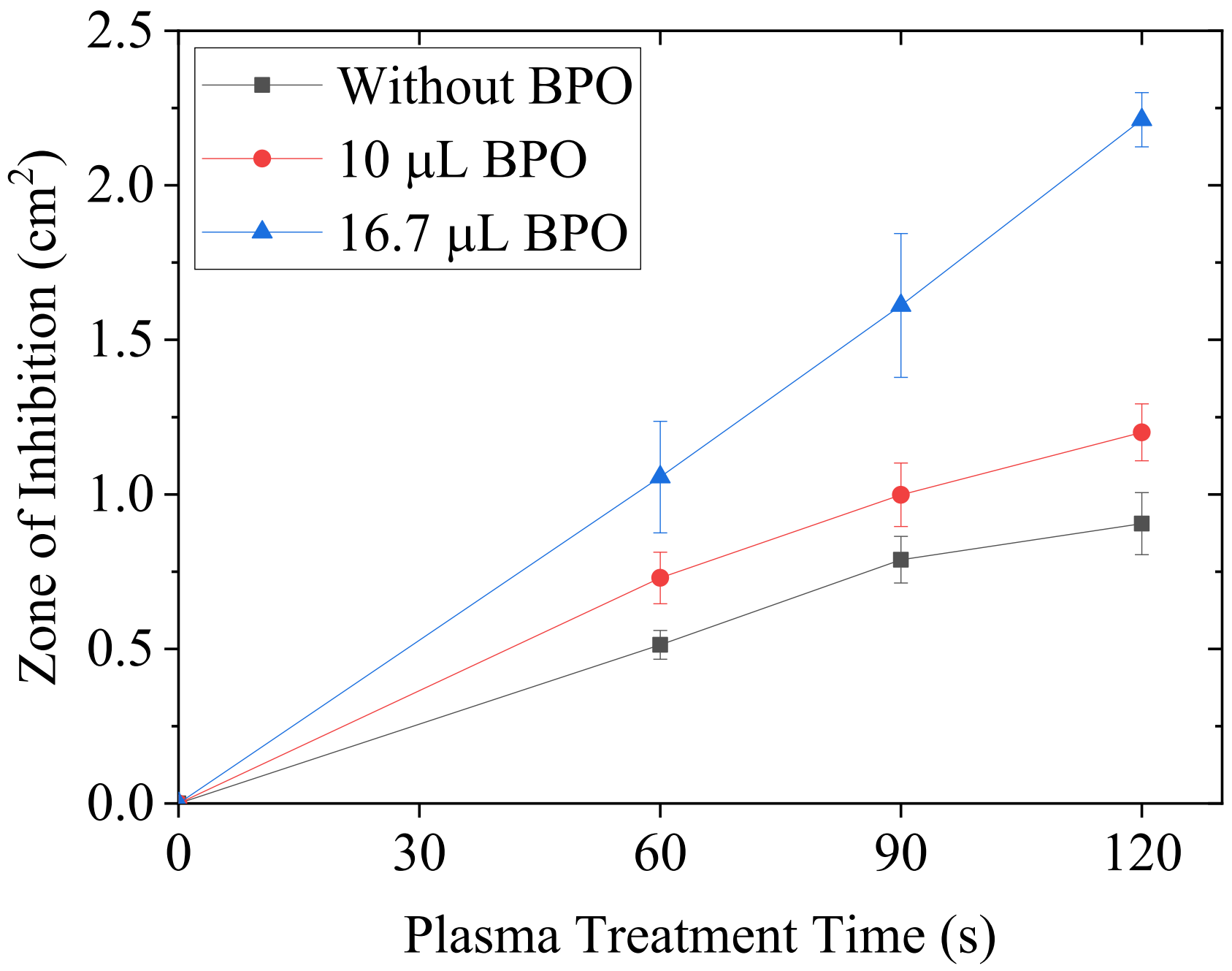

**FIG. 6:** The effect of BPO addition on the plasma sterilization area against *E. coli*.

Figure 6 shows the effect of plasma and BPO on the sterilization area of *E. coli*, with the plasma treatment distance fixed at 17 mm. The zones of inhibition without BPO at 60, 90, and 120 seconds of plasma treatment was 0.51 ± 0.047, 0.8 ± 0.075, and 0.91 ± 0.100 cm², respectively. When 10 μL of BPO was added, the zones of inhibition increased to 0.73 ± 0.083, 0.99 ± 0.103, and 1.2 ± 0.092 cm², respectively. With the addition of 16.7 μL of BPO, the zones of inhibition further increased to 1.6 ± 0.180, 1.61 ± 0.233, and 2.21 ± 0.088 cm², respectively.

Based on the sterilization area measurement experiments for *E. coli*, under the same plasma exposure time, the sterilization areas were the smallest when no BPO was added. However, after the addition of BPO solution, the sterilization areas were significantly increased, and the more BPO solution added, the better the enhancement effect. This indicates that the BPO solution can indeed assist plasma in achieving better sterilization efficacy.

The thickness of the *E. coli* layer surrounding the zones of inhibition decreased progressively across the three groups, no BPO, 10 µL BPO, and 16.7 µL BPO, likely reflecting the effect of BPO.

Due to the relatively long plasma treatment distance of 17 mm adopted in this study, the plasma exposure became less focused, resulting in an irregular sterilization pattern rather than a well-defined circular area.

Compared with either treatment alone, the enhanced efficacy observed with the combined BPO and plasma treatment may be attributed to:

1. Plasma may enhance the decomposition of BPO, thereby increasing its bactericidal effect.
2. Plasma may disrupt the bacterial envelope, weakening bacterial defense mechanisms and making them more susceptible to BPO, or facilitating the penetration of BPO into the cells to exert its toxic effects.
3. The simultaneous action of ROS and BPO may more rapidly drive bacteria to an irreversible state of damage, resulting in a more pronounced bactericidal effect.
4. Interactions between plasma-generated ROS and BPO may produce additional substances that are harmful to bacteria.

Based on the current results, we are not yet able to clarify the major mechanism. The UV emission from plasma may be insufficient to induce significant BPO decomposition in our system. The observed effects can be more likely attributable to synergistic interactions between UV radiation and plasma-generated reactive species, which collectively enhance BPO activation and subsequent biological activity.

However, previous studies on plasma treatment in cellular systems have shown that plasma can, in many cases, effectively enhance drug uptake,[20-22] and the ROS threshold may also be one of the primary mechanisms in cancer cell therapy.[23] Therefore, we consider possibilities (2) and (3) to be more plausible. The exact mechanism requires further experimental validation.

The results of the antibacterial assay indicate that the combination of BPO with cold atmospheric pressure plasma significantly enhances bactericidal efficacy compared with plasma or BPO treatment alone, suggesting strong potential for practical applications. Although the precise chemical interactions between plasma-generated species and BPO remain to be fully elucidated, the observed enhancement in inhibition zones supports the feasibility of this combined approach as an effective and clinically relevant antibacterial strategy.

# IV. Conclusion

In this study, the optimal plasma parameters for the experimental setup were determined through OES, with parameters set to 4.3 kV for voltage, 4 SLM for argon gas flow rate, and no oxygen addition. Surface temperature measurements and UV emission analysis were conducted to characterize the thermal and optical properties of the plasma treatment. Although chemical titration results did not provide conclusive evidence for the decomposition of BPO by plasma, it is suggested that any possible interaction is more likely driven by plasma-generated reactive species rather than ultraviolet radiation. The practical sterilization experiments confirmed that the addition of BPO effectively enhanced plasma's bactericidal activity against *E. coli* (0.91 $cm^2$ → 2.21 $cm^2$ after 120 seconds of plasma treatment). This synergistic effect highlights the potential of combining chemical agents with plasma to improve antibacterial performance. Future studies should further investigate the underlying mechanisms, plasma chemistry, biocompatibility, and optimization of device configurations for clinical translation in biomedical, dental, and dermatological applications.

## Acknowledgements

The authors would like to thank Prof. Jong-Shinn Wu from National Yang Ming Chiao Tung University (NYCU) for providing the plasma equipment. Part of this study was previously presented at the National Conference on Mechanical Engineering of CSME, Taiwan, in 2024.

# Appendix

## I. Zone of Inhibition Against *E. coli*

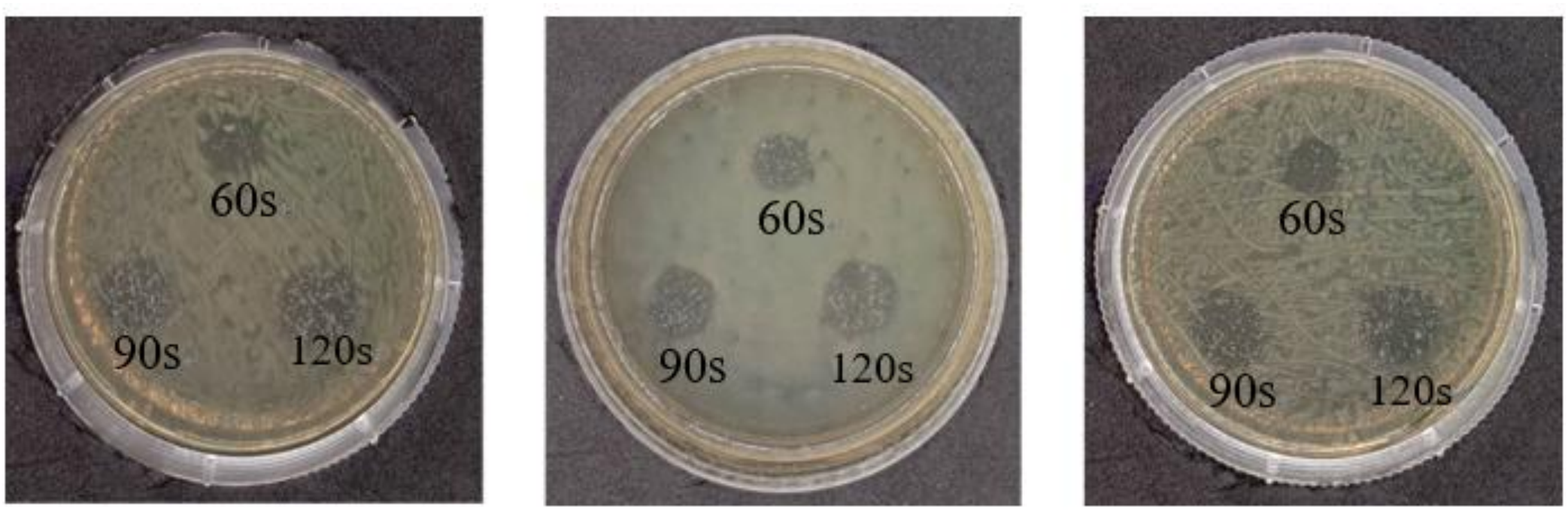

**FIG. 1:** Plasma sterilization area against *E. coli* with triplicates.

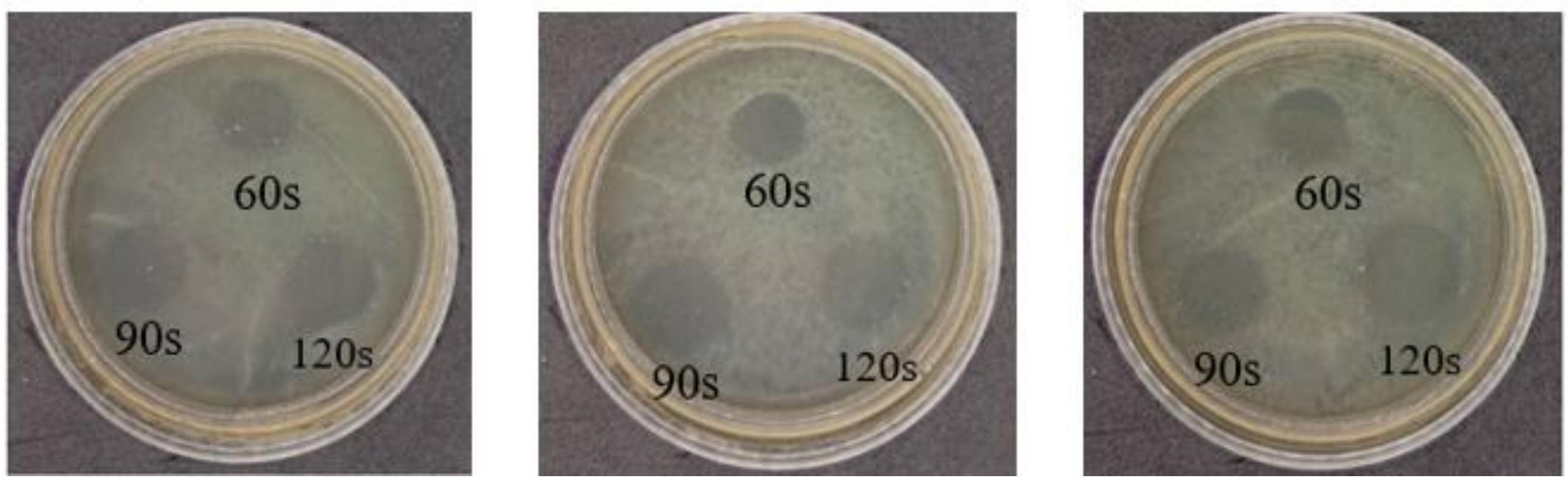

**FIG. 2:** Plasma with the addition of 10 μL BPO on the sterilization area against *E. coli* with triplicates.

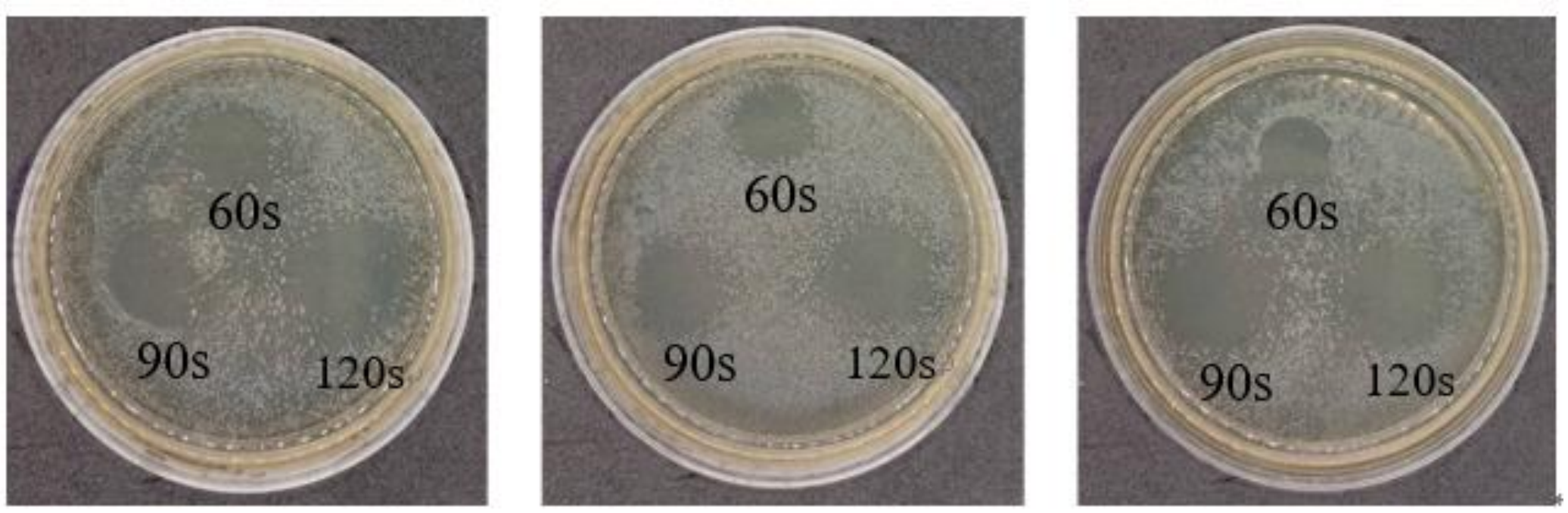

**FIG. 3:** Plasma with the addition of 16.7 μL BPO on the sterilization area against *E. coli* with triplicates.

## II. Surface Temperature

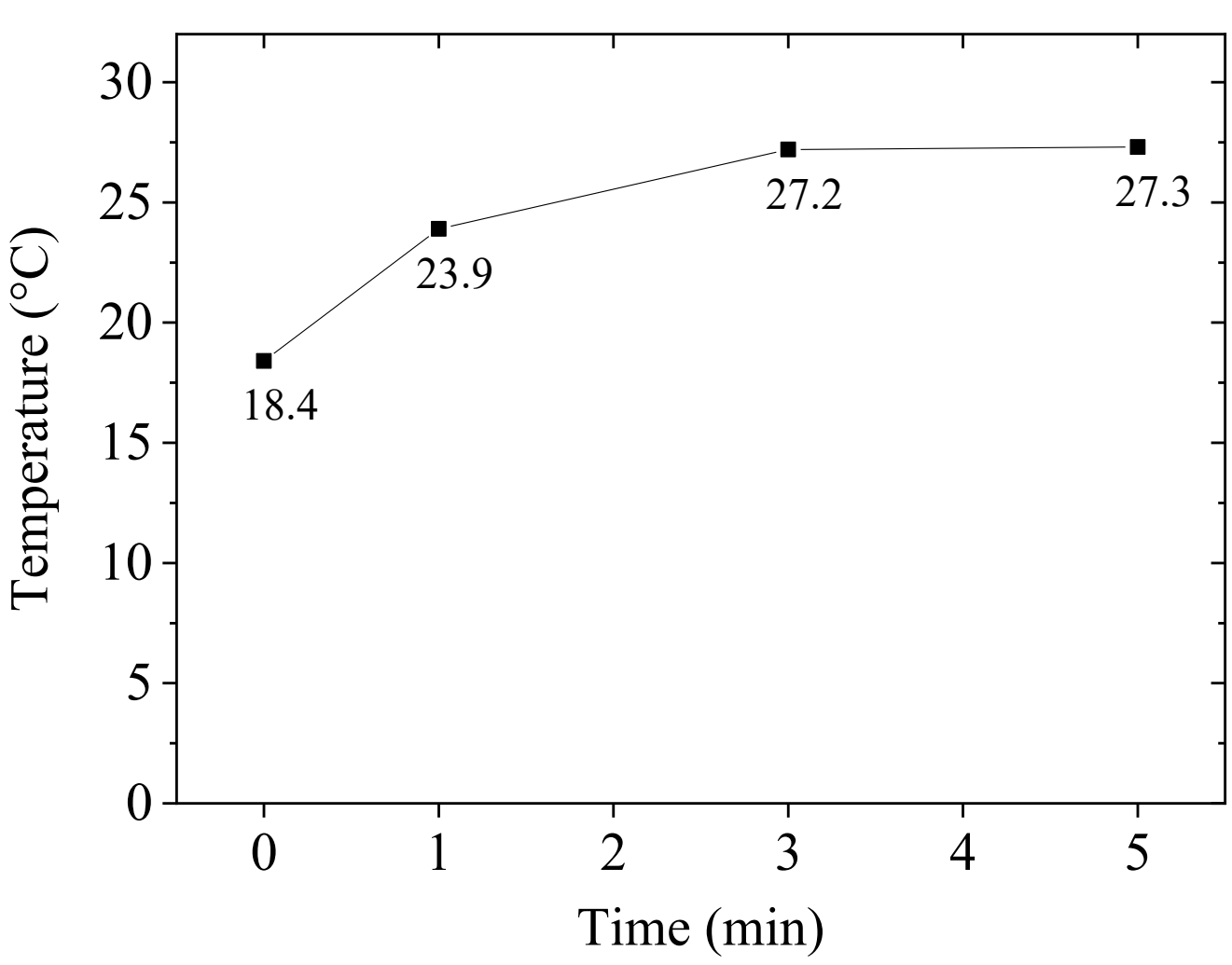

**FIG. 4:** Surface temperature measurement of ceramic material using FLIR for varied treatment durations.